\documentclass[prl,twocolumn,aps,superscriptaddress,preprintnumbers,nofootinbib]{revtex4}
\usepackage{latexsym}
\usepackage{graphicx}
\usepackage{amsfonts}
\usepackage{amssymb}
\usepackage{amsmath}
\usepackage{hyperref}
\usepackage{slashed}
\usepackage{empheq}
\usepackage{multirow}
\usepackage{fancyhdr}
\usepackage{appendix}
\usepackage{subfigure}
\usepackage{times}

\usepackage{graphics,graphicx}
\usepackage{amsmath,amsfonts,amssymb,indentfirst,mathrsfs,bm,fancyhdr}
\usepackage[utf8]{inputenc}
\usepackage[T1]{fontenc}
\usepackage{esint}
\usepackage{hyperref}
\usepackage{color}
\usepackage{float}

\newcommand{\be}{\begin{equation}}
\newcommand{\ee}{\end{equation}}

\def\bea{\begin{eqnarray}}
\def\eea{\end{eqnarray}}

\def\beqx{\begin{displaymath}}
\def\eeqx{\end{displaymath}}

\newcommand{\bmat}{\left(\begin{array}}
\newcommand{\emat}{\end{array}\right)}

\def\bo{{\raise-.3ex\hbox{\large$\Box$}}}               
\def\face{{\raise.2ex\hbox{$\displaystyle \bigodot$}\mskip-2.2mu \llap {$\ddot
        \smile$}}}                                   
\def\>{\rangle}                                      
\def\<{\langle}                                      

\def\leftrightarrowfill{$\mathsurround=0pt \mathord\leftarrow \mkern-6mu
        \cleaders\hbox{$\mkern-2mu \mathord- \mkern-2mu$}\hfill
        \mkern-6mu \mathord\rightarrow$}        
\def\dvec#1{\vbox{\ialign{##\crcr
        \leftrightarrowfill\crcr\noalign{\kern-1pt\nointerlineskip}
        $\hfil\displaystyle{#1}\hfil$\crcr}}}           

\def\-{\hphantom{-}}

\begin{document}

\preprint{Imperial-TP-LM-2014-01}

\title{Anomalous massless modes}

\author{Luis Melgar}\email{luis.melgar@imperial.ac.uk}
\affiliation{Imperial College London,\\ Blackett Laboratory, SW7 2AZ,\\London, U.K.}

\begin{abstract} 
Some years ago Anton Yu. Alekseev \emph{et al.} conjectured the existence of massless modes in the spectrum of excitations (``anomalous massless modes'') building upon certain similarities between a spontaneous symmetry breaking and the interplay of axial and vector symmetries in an anomalous theory. We reinterpret the analogy and argue that the presence of these modes is ensured in any (even) number of dimensions only by the excitation of certain anomaly-induced terms called Schwinger terms. In 1+1 dimensions the anomalous massless mode corresponds to the charge density wave present in a Luttinger liquid. In 3+1 dimensions, we identify the anomalous massless mode with the so-called chiral magnetic wave. Our analysis shows that both modes arise as a consequence of the same physics.
\end{abstract}

\pacs{ }

\maketitle

\section{Introduction}
\label{sec:intro}
It is well known that anomalies play a very important role in the definition and consistency of Quantum Field Theories (QFT's) \cite{Bertlmann:1996xk}. For instance, anomalies in the gauge sector must be avoided since they spoil the renormalizability of the theory.\\
Tightly related to anomalies are the so-called Schwinger terms (ST), which are terms appearing in the equal time commutators of constrain operators \cite{Faddeev:1984jp}, inducing an extension of the algebra that can be observed at the level of current commutators \cite{PhysRev1821459}. Like anomalies, ST can be obtained by means of the Stora chain of descent equations and known results indicate that they are robust against interactions \cite{Sykora:1999xw}. 
Schwinger terms have also implications for theories at finite temperature and/or chemical potential. 
This fact is eventually responsible for the chiral magnetic effect \cite{Fukushima:2008xe}. As we will briefly review later on, one can also extract the chiral magnetic effect from the existence of ST. Associating anomalous transport to ST is more natural than using the anomaly polynomial. The reason is that the chiral magnetic effect is present even when the anomaly is not excited, but Schwinger terms need to be switched on for the chiral magnetic effect to be present, as we will see. 
\\

Focusing on the interplay between a vector and axial symmetry, the purpose of this note is to emphasize that an anomaly (concretely, the presence of ST) leads to the existence of massless modes in the spectrum of excitations  (``anomalous massless modes'') and to analyze some of the consequences of their presence. To that end, we start out motivating the idea in the next section by studying the 1+1-dimensional case in detail. Then we extend the result for $D=d+1$-dimensional constructions (with $D$ even), with similar conclusions to the ones of \cite{antonyuetal}; we elaborate on the main differences between both approaches later on. Then we move on to the particular case of 3+1 dimensions and interpret the resulting anomalous massless mode as the so-called chiral magnetic wave (CMW) \cite{Kharzeev:2010gd}, which has been conjectured to have implications for heavy-ion collisions 
The whole construction shows certain striking similarities with the physics of spontaneous symmetry breaking, even though, as will become clear later on, there is no charged operator developing a vacuum expectation value in our setup. Our construction makes the presence of CMW a general and robust feature, as long as there exists an anomalous symmetry plus a background magnetic field. Let us remark that Schwinger terms are indeed crucial to make the connection; in its absence our conclusions are not valid anymore. We conclude elaborating on future directions and possible generalizations based on the known results regarding usual NG bosons.\\
It is important to clarify that throughout this paper we will be considering \emph{non-dynamical} external gauge fields, so that we can consider a background in which the divergence of the current is effectively conserved. 
\section{Fermion in 1+1 dimensions, Schwinger Term and Luttinger liquids}
\label{sec:1+1}
Let us discuss a general system of left and right-moving fermions in 1+1 dimensions $\psi_L$ and $\psi_R$, with the following lagrangian
\begin{align}
\mathcal{L}= \bar \psi_L i \slashed{\partial} \psi_L +\bar \psi_R i \slashed{\partial} \psi_R + \mathcal{L}_{\text{int}}  \,,
\end{align}
where $\mathcal{L}_{\text{int}}$ represent a generic interacting term, which should not affect cualitatively our forecoming conclusions, due to the robustness of ST.\\
It is very useful to perform the usual bosonization 
\begin{align}
\psi_L=e^{- \pi i \phi_L};\ \ \psi_R= e^{ \pi i \phi_R}\,,
\end{align}
where $\phi_{L,R}$ is a scalar field. 
Moreover, we further split $\phi_R = \frac{1}{2}\left(\varphi +v\right)$ and $\phi_L = \frac{1}{2}\left(\varphi -v\right)$. It turns out that $\varphi$ is \emph{dual} to $v$, i.e. $\partial_t \varphi= -\partial_x v$ (see for instance \cite{1999Senechal}). This leads us to an expression for vector current can be written in terms of the scalar $\varphi$ as
\begin{align}
\label{eq:dual1+1}j^{\mu}_v = \epsilon^{\mu \nu} \partial_{\nu} \varphi\,,
\end{align}
which in particular implies that the charge density is $j^0_v = \psi_L^{ \dagger} \psi_L +\psi_R^{ \dagger} \psi_R = \partial_x \varphi$. Moreover, exploiting the identity between gamma matrices in two dimensions $\gamma_{\mu} \gamma_5= \epsilon_{\mu \nu}\gamma^{\nu}$ we can identify $j^{\mu}_a (x) = \partial^{\mu} \varphi(x)$. Now,  the Schwinger term provides us with a central extension, and the (anomalous) commutator of the axial and vector charge densities reads \cite{Adam:1997hs}
\begin{align}
\label{eq:ST1+1} \left[j^{0}_a(x), j^0_v(y)\right] = \frac{i}{2 \pi} \partial_1 \delta(x-y)\,,
\end{align}
or
\begin{align}
\label{eq:STs1+1} \left[Q_a,\varphi\right] = \frac{i}{2 \pi} \,,
\end{align}
where we have defined the charge operator $Q_a = \int dx_1\ j^0_a(x_0, x_1)$. The anomaly-induced commutator (\ref{eq:STs1+1}) has important consequences. Equation (\ref{eq:STs1+1}) indicates that $\varphi$ gets shifted under the action of $Q_a$. Indeed, we claim that the scalar field $\varphi(x)$ is the analogue of a goldstone field for a broken $U(1)$ symmetry.
As expected, a constant shift of $\varphi$ does not make any difference (in particular, our currents are invariant under it), but if we assume that $\varphi(x) = \varphi_0 + \varphi_1(x)$, with the latter being a small perturbation, we find that $\varphi_1(x)$ represents a massless field. In other words, preserving the axial symmetry implies that the effective action must be invariant under constant shifts of $\varphi$ and thus cannot contain any mass-term for that scalar\footnote{We thank Carlos Hoyos for suggesting this argument.}. As a consequence, there must be a massless mode in the spectrum of excitations of the theory, namely, our anomalous massless mode. \\
Notice in passing that at finite chemical potential, we have $j^1_v = \mu_a/4\pi$, which is nothing but the chiral magnetic effect in 1+1 dimensions. But since $j^1_v= \partial_t \varphi(x)$, we conclude that $\partial_t \varphi(x) =4 \pi \mu_a$, which is, up to a trivial factor, the usual relation between the (superfluid) chemical potential and a derivative of a goldstone field (this is also compatible with $j^{\mu}_a (x) = \partial^{\mu} \varphi(x)$). This is another example of the similarities between the anomalous massless mode and a true Nambu-Goldstone mode.\\

\paragraph{Anomalous massless boson from hydrodynamics:} As expected, a new massless mode can be observed by a hydrodynamic analysis of the theory. As will become clear, it turns out that the presence of these massless modes is related to the alterations in the hydrodynamics that are induced by the so-called anomalous transport in the corresponding dimension.\\
Indeed, the currents feature the following hydrodynamic expansion (we have rescaled the chemical potentials to avoid factors of $4\pi$ in order to make the discussion clearer)
\begin{align}
\nonumber j^{1}_v =  \mu_5 -  D_v \partial_{x^1} j^{0}_v +\mathcal{O}(k^2) \,,\\
j^{1}_a =  \mu -  D_a \partial_{x^1} j^{0}_a+\mathcal{O}(k^2)\,,
\end{align}
where $\{v,a\}$ stand for ``axial'' and ``vector'', $D_{\{v,a\}}$ are diffusion constants and $\mu_5 = j^0_a/\chi_a, \mu=j^0_v/\chi_v$ are axial and vector chemical potentials respectively. Since the electric field vanishes, we can impose current conservation to obtain
\begin{align}
\nonumber \omega j^0_v + j^0_a\frac{k }{\chi_a} + i k^2 D_v j^0_v =0\,,\\
\omega j^0_a + j^0_v \frac{k }{\chi_v} + i k^2 D_a j^0_a =0\,,
\end{align}
leading to dispersion relations
\begin{align}
\label{eq:disp}\omega_{\pm}(k) = \pm \frac{1 }{\sqrt{\chi_a \chi_v}}k - \frac{i}{2} (D_a+D_v) k^2 + \mathcal{O}(k^3)\,.
\end{align}
So we observe a massless mode with a linear dispersion relation. Although not explicit, it is important to point out that its velocity of sound would vanish in the absence of the ST, which induces the central extension in (\ref{eq:STs1+1}).
\subsection{Relationship with Luttinger liquids}
\label{subsec:Luttinger}
As will be shown later on, in general ST depend on the presence of background gauge fields that we can turn off at will. However, according to (\ref{eq:STs1+1}), this is not the case in 1+1 dimensions. Therefore, any consistent theory of fermions in 1+1 dimensions should take this issue into account. This was initially done in \cite{MattisLieb}, in which the authors provided a solution to the Luttinger liquid via bosonization\footnote{Equation (\ref{eq:STs1+1}) is crucial to arrive at this intuition because it can be interpreted as the commutation relation of a scalar field and its conjugate momentum.}. In this theory one can show that the bosonization of the Luttinger liquid leads to the following action for our bosonization field $\varphi(x)$ (see for instance \cite{Giamarchi:743140})
\begin{align}
\label{eq:SB}S_{\varphi}\propto \int d^2 x\left[ \left(\partial_0 \varphi\right)^2 - v^2 \left(\partial_1 \varphi\right)^2\right]\,,
\end{align}
that is, field appearing in equation (\ref{eq:STs1+1}) leads to a wave-like massless mode with velocity of sound $v$ (which is fixed by the couplings of the original fermionic theory).\\
 In the context of Luttinger liquids this mode is called charge density wave. We therefore reach the conclusion that our anomalous massless mode in 1+1 dimensions is nothing but the well-known charge density wave arising in Luttinger liquids. \\
 The connection with Luttinger liquids is interesting for another reason. It allows us to elucidate to what extent the field $\varphi(x)$ is the goldstone field arising from a \emph{real} spontaneous symmetry breaking. 
In particular, notice that we can construct an operator charged under $Q_a$ of the form
\begin{align}
\label{eq:opar1+1}\mathcal{O} =\psi^*_L \psi_R = e^{\pi i \varphi}\,,
\end{align}
where we have used the symmetry $\varphi \rightarrow \varphi + \varphi_0$. The question we want to address is then \emph{does $\mathcal{O}$ acquire a vacuum expectation value (VEV)?} The answer is obviously negative due to the Mermin-Wagner theorem. Making use of the Luttinger-liquid theory we can be more explicit. The two-point function $\mathcal{G}(x) = \<\mathcal{O}(x)\mathcal{O}(0)\>$ behaves at large distances as \cite{Giamarchi:743140}
\begin{align}
\label{eq:corrOO} \mathcal{G}(x>>1) \sim \left(\frac{1}{x}\right)^{\gamma} +...\,,
\end{align}
where $\gamma>0$. We see that  $\mathcal{G}(x)$ decays at long distances, indicating that there is no long-range order. We conclude that
\begin{align}
\label{eq:VEV1+1} \<\mathcal{O}\>  =0\,,
\end{align}
and hence, as expected, although there exists an anomalous massless mode that shares many features with a NG boson, there is no real spontaneus symmetry breaking in our system. 
\section{Anomalous massless boson in general (even) dimensions}
\label{sec:chiralD}
Our considerations in the previous section generalize without obstruction to higher (even) dimensions and thus we expect the same conclusions to hold. However, in $D=d+1>2$ dimensions there are several technical differences that arise and we proceed to point out.\\
First of all, the central extension (\ref{eq:ST1+1}) becomes more complicated in dimension $D>2$, and generically depends on external fields $\mathcal{F}$, schematically
\begin{align}
\label{eq:STD}\left[j^0_a(x), j^0_e(y)\right]  = i f(\mathcal{F})^i \partial_i\delta(x-y)\,,
\end{align}
where $i= 1,...,d$ and $f$ is a (vector) functional of the external fields. In general, we should consider background fields such that $f(\mathcal{F})^i \ne 0$, which turns the ST on. In 1+1 dimensions this is not a problem because $f$ is a constant in that case. Moreover, there exist a further constraint on our background configuration, namely, it has to be such that the anomaly is switched off. We will see later on that it is not complicated to come up with a background configuration such that all the requirements are satisfied.\\

Apart from the background configuration, there is another complication for $D>2$ that has to do with the extension of (\ref{eq:dual1+1}); one way to proceed is to introduce an off-shell and ad-hoc solution to the conservation equation $\partial_{\mu}j^{\mu}_v=0$ in terms of a $d-1$-form $\Lambda$, as 
\begin{align}
\label{eq:currD}j^{\mu}_v  =\epsilon^{\mu \mu_1...\mu_d} \partial_{\mu_1}\Lambda_{\mu_2...\mu_d}\,.
\end{align}
However, there is some gauge freedom $\Lambda\rightarrow \Lambda + dg$ remaining. After gauge-fixing, we are left with a $d-$dimensional vector $\vec \omega$. This is the procedure employed in \cite{antonyuetal} in order to find the analogue of the scalar field $\varphi$ for $D>2$. This procedure finds more justification when trying to generalize the bosonization procedure to dimensions higher than $D=2$. One can see that one is then naturally lead to a Kalb-Ramon gauge potential that can be identified with $\Lambda$ \cite{Burgess:1994tm, Frohlich:1994mj}, which in turn provides us with a current of the form (\ref{eq:currD}) (see for instance \cite{Schaposnik:1995np}). Therefore, we take the the $d$-dimensional vector field $\omega_j(x)$ as the proper generalization of $\varphi(x)$ (observe that the 1+1-dimensional results are recovered if we apply this general logic back to the case $D=2$).\\

Equipped with the above considerations, we can now apply the conclusions reached in the 1+1-dimensional case. The field $\omega_j$ fulfills 
\begin{align}
\label{eq:STwD}\left[Q_a, \omega_j (y)\right]  = i f(\mathcal{F})_j(y) \,,
\end{align}
and similarities with a goldstone field are again manifest. For the same reasons as in $D=2$, field $\omega_j$ will induce a massless excitation, which is an anomalous massless boson in $D$ dimensions. Observe that, once again, Schwinger terms (\ref{eq:STD}) or (\ref{eq:STwD}) are crucial to induce non-trivial transformation properties of $\omega_j$ under an axial transformation; there will be no gapless mode in its absence.\\
A similar conclusion was reached in \cite{antonyuetal}. In that reference the VEV of the commutator in (\ref{eq:STwD}) is interpreted as some sort of order parameter and Goldstone's theorem is invoked to argue in favor of the existence of massless modes whenever the current satisfies $\<j^{i}_v\> \ne 0$. Here, on the contrary, we do not rely on Goldstone's theorem but only on the non-trivial transformation of $\omega_i$ under $Q_a$ at the operator level\footnote{In fact, we have seen that $\omega_i$ (or $\varphi$) can be interpreted more naturally as goldstone fields, not as some order parameter operator. As shown explicitly in $D=2$, the VEV of the charged operator $\mathcal{O}$ vanishes.}. This allows us to link the existence of massless modes to the presence of non-vanishing Schwinger terms in the corresponding dimension, without appealing to any other quantity.\\
In the following, we particularize the above general discussion to the $3+1$-dimensional case. 
\subsection{Anomalous massless boson in 3+1 dimensions and the Chiral Magnetic Wave}
\label{subsec:3+1}
To illustrate the previous comments, we focus now on the 3+1-dimensional case. We assume that we have a chiral fermion and consider the interplay between axial and vector currents. The Schwinger term now takes the form \cite{Treiman:1986ep}
\begin{align}
\label{eq:Schwing3+1}\left[j^0_a(x), j^0_v(y)\right] = i \lambda B_j(x) \partial_j \delta(x-y)\,,
\end{align}
which does not vanish, but depends on a external magnetic field $B_j(x)$ (with $j=1,2,3$). The coefficient $\lambda$ above is related to the anomaly coefficient. The ST can be recast in the more convenient form\footnote{Rotation invariance allows us to choose a given direction of the magnetic field, so practically speaking, only one of the components of $\vec \omega$ will actually lead to a non-trivial commutator with $Q_a$.}
\begin{align}
\label{eq:gammaSchwing3+1}\left[Q_a, \omega_j\right] = i \lambda B_j\,.
\end{align} 
In order to ensure the conservation of the currents we are required to assume that our background configuration is such that the electric field vanishes $\vec E=0$, whereas the magnetic field is non-zero $\vec B\ne 0$\footnote{Actually, $\vec E \ne0$ is also allowed as long as $\vec E\cdot \vec B =0$.}.

At this point it is worth mentioning the result of \cite{antonyuetal}, in which Schwinger term (\ref{eq:gammaSchwing3+1}) is used to derive the chiral magnetic effect. The authors consider a finite chemical potential, and relate equation $\vec j_v(x) =- \partial_t \vec \omega$ to the ST (\ref{eq:gammaSchwing3+1}). 
This provides the result\footnote{At this point, it is important to make sure that one is working with a suitable definition of the current such that the chemical potential $\mu_5$ is well defined and equation (\ref{eq:CME3+1}) holds.}
\begin{align}
\label{eq:CME3+1}\<\vec j_v\> (x)= -\frac{i}{2 \hbar}\mu_5 \int dy\<\left[j^0_a(y), \vec \omega(x)\right]\> = \frac{\lambda}{2 \hbar} \mu_5 \vec B(x)\,,
\end{align}
which in particular is valid for arbitrary (namely, non-necessarily constant) magnetic field. In this way, we can derive the chiral magnetic effect without any reference to the anomalous divergence of the current in the presence of external fields, which is more natural for we have been working all the time with $\vec E=0$, so that the anomaly $\vec E\cdot \vec B=0$. This construction again emphasizes the tight relation between the presence of anomalous massless bosons and the existence of anomalous transport. However, notice an important difference: equation (\ref{eq:CME3+1}) vanishes if $\mu_5=0$, whereas the existence of anomalous gapless bosons is ensured even at zero chemical potential by arguments given above, as long as the magnetic field is present. \\
 
 Applying the same logic as in previous sections, we thus expect an anomalous massless mode in the spectrum of excitations. As before, we assume that temperature is different from zero and use a hydrodynamic approach. In fact, the computation is identical to that carried out in 1+1 dimensions, the only difference being that now the external magnetic field enters explicitly the constitutive relations for the spatial components of the currents.
Assuming that $\vec B= B\vec e_x$ points in the $x$-direction, the hydrodynamic expansion of the currents now reads
\begin{align}
\nonumber j^x_v = C \mu_5 B -  D_v \partial_x j^{0}_v +...\,,\\
\nonumber  j^x_a = C \mu  B-  D_a \partial_x j^{0}_a+...\,,
\end{align}
where $C\equiv \frac{\lambda}{2 \hbar}$ is (proportional to) the anomaly coefficient. Imposing conservation of the above currents leads to the following dispersion relations ($k\equiv k^x$)
\begin{align}
\label{eq:disp3+1}\omega_{\pm}(k) = \pm \frac{C B}{\sqrt{\chi_a \chi_v}}k - \frac{i}{2} (D_a+D_v) k^2 + \mathcal{O}(k^3)\,.
\end{align}
As aforementioned, our anomalous massless mode in 3+1 dimensions is the chiral magnetic wave \cite{Kharzeev:2010gd}, which has been found recently as a consequence of anomalies. Our analysis shows that the presence of this mode is completely general as long as we have an AVV anomaly and a background magnetic field.\\
To conclude this section, let us remark again that the presence of this mode is independent of the chemical potential. This feature is somehow obscured by the hydrodynamic derivation, but one can use holographic methods and encounter the anomalous massless mode (the CMW in 3+1 dimensions) in the spectrum of a thermal theory even at zero chemical potential (notice that the chemical potential does not appear in (\ref{eq:disp3+1})).
\section{Conclusions and future directions}
\label{sec:Concs}
In this note we have shown that, as long as Schwinger terms are excited, anomalies induce massless modes in the spectrum of excitations (``anomalous massless modes''). This is compatible with the fact that they alter the hydrodynamics of the theory (giving rise to the chiral magnetic and chiral vortical effects). \\
We have investigated how a Schwinger term in 1+1 dimensions implies the existence of a massless mode. We also pointed out the connection with Luttinger liquids; the anomalous gapless boson gets in this way identified with the charge density wave. Besides, there is no real spontaneous symmetry breaking in our setup. 
Then, we have generalized the above logic to higher dimensions and worked out the 3+1-dimensional case as well. In 3+1 dimensions the anomalous massless boson corresponds to the chiral magnetic wave \cite{Kharzeev:2010gd}, which arises naturally and in an expected way in our framework. In this way, we unify the physics of the CMW with that of the charge density wave. Our conclusions imply the presence of massless modes for arbitrary (even) dimensions as long as Schwinger terms are excited.\\

We expect a rich phenomenology associated to anomalous massless bosons. For instance, they should have a great impact on the thermodynamic properties of the system. Moreover, it is well-known that the CMW provokes an infinite D.C. conductivity (this is known under the name of negative magnetoresistivity \cite{Nielsen:1983rb} which is expected to play a role in Dirac and Weyl-semimetals \cite{Son:2012bg,Gorbar:2013dha}), very much as a usual NG boson does. It would be interesting to study whether this feature is independent of dimensionality. Exploiting further the analogy, let us also point out that the presence of a vector chemical potential switches on the spatial component of the axial current, and hence it is possible that the anomalous gapless boson behaves in the presence of a finite vector chemical potential as a usual NG boson does whenever we turn on supervelocity. If this is so, there would be an associated Landau criterion and a corresponding instability for large enough vector chemical potential. \\
In fact, due to the similarities between anomalous massless modes and usual NG bosons, it is plausible that conclusions in the analysis of the latter carry over to our case. If so, an interesting extension would be to address the possibility of constructing a theory featuring Type-II anomalous gapless bosons (there is a lot of literature on usual Type-II NG bosons \cite{Nielsen:1975hm}, some of it quite recent; see \cite{Watanabe:2012hr} and also \cite{Brauner:2010wm} for a review), which typically involves either explicit or spontaneous breaking of Lorentz symmetry to some extent, as well as the interplay between several generators of a non-abelian symmetry. Furthermore, for relativistic theories it is known that Type-II NG bosons are accompanied by certain ``massive goldstones'' \cite{Nicolis:2012vf,Watanabe:2013uya}. 

There are other future directions that one can follow, regarding explicit set-ups in which anomalous massless bosons can be observed. For instance, there can exist further anomalous sound modes if Schwinger terms affect the commutator of a current and the energy density (possibly, such terms would be proportional to the external vorticity). Another direction involves the extension of our arguments to Weyl and gravitational anomalies. However, in that case it is not straightforward to establish a physical effect because, as shown in \cite{Bertlmann:2000da} for the 1+1-dimensional case,
\begin{align}
\label{eq:Weylgravn}  \<\left[ T_a(x), T_v(y)\right]\> \propto \partial_1^3\delta(x-y)\,,
\end{align}
where $T_{\{a,v\}}$ denotes relevant components of the axial/vector energy-momentum tensor, is of higher order in derivatives. If we assume that the energy-momentum tensor can be written in terms of a putative scalar field as $T \sim \partial^2 \theta$, the derivative that is left in (\ref{eq:Weylgravn}) invalidates the program carried out above. But a cautionary remark is in order here. The naive momentum power counting is known to led to wrong conclusions when analyzing physical effects associated to the mixed gauge-gravitational anomaly \cite{Landsteiner:2011cp, Jensen:2012kj}.\\
Finally, our argument could break down if we excite the anomaly in the background. This would be consistent with considerations presented in \cite{Jimenez-Alba:2014iia} and could be investigated further.   
\section*{Acknowledgments}
I would like to thank Carlos Hoyos, Karl Landsteiner, Amadeo Jimenez-Alba and Natalia Pinzani-Fokeeva for fruitful discussions and comments on the draft. The work of L.M. was supported by the ERC Advanced grant No.339140 ``Gravity, Black Holes and Strongly Coupled Quantum Matter''.
\addcontentsline{toc}{section}{References}

\nocite{*}

\bibliographystyle{jhepcap}
\bibliography{Angaplessbib2}

\end{document}